# Quantum Multiplier Based on Exponent Adder

Junpeng Zhan[1]

[1]Department of Renewable Energy Engineering, Alfred University, Alfred, New York 14802, USA

zhanj@alfred.edu

(Dated: April 4, 2024)

## Abstract

Quantum multiplication is a fundamental operation in quantum computing. It is important to have a quantum multiplier with low complexity. In this paper, we propose the Quantum Multiplier Based on Exponent Adder (QMbead), a new approach that requires just $\log_2(n)$ qubits to multiply two $n$-bit integer numbers, in addition to $O(n)$ ancillary qubits used for quantum state preparation. The QMbead uses a so-called exponent encoding to respectively represent two multiplicands as two superposition states which are prepared by a *quantum state preparation* method, then employs a *quantum adder* to obtain the sum of these two superposition states, and subsequently measures the outputs of the quantum adder to calculate the product of the multiplicands. Different quantum adders can be used in the QMbead. The *circuit depth* and *time complexity* of the QMbead, using a logarithmic-depth quantum carry lookahead adder (QCLA) as adder, are $O(\log n)$ and $O(n \log n)$, respectively. The *gate complexity* of the QMbead is $O(n)$. The circuit depth and gate complexity of the QMbead is better than existing quantum multipliers such as the quantum Karatsuba multiplier and the QFT based multiplier. The time complexity of the QMbead is identical to that of the fastest classical multiplication algorithm, Harvey-Hoeven algorithm. Interestingly, the QMbead maintains an advantage over the Harvey-Hoeven algorithm, given that the latter is only suitable for excessively large numbers, whereas the QMbead is valid for both small and large numbers. The multiplicand can be either an integer or a decimal number. The QMbead has been successfully implemented on quantum simulators to compute products with a bit length of up to 273 bits using only 17 qubits, excluding the ancillary qubits used for quantum state preparation. This establishes QMbead as an efficient solution for multiplying large integer or decimal numbers with many bits.

## I. Introduction

Quantum computing, with its potential for exponential speedup in solving complex problems, has revolutionized the landscape of computational algorithms. In the domain of quantum computing, the pursuit for highly efficient quantum arithmetic algorithms is of paramount importance. Shor's algorithm [1], [2] demonstrated an exponential advantage of quantum computing in factoring large numbers, a problem crucial to cryptography. This breakthrough has motivated extensive research and development efforts aimed at advancing quantum arithmetic, including, but not limited to, adders [3], [4], [5], [6], modular adder [5], [6], [7], [8], multipliers [6], [9], [10], [11], [12], [13], [14], modular multiplication [8], [9], [15], matrix multiplication [16], [17], [18], [19], [20], and modular exponentiation [5], [8], [21].

This paper focuses on quantum multipliers. A common and straightforward approach to construct a quantum multiplier is through the repetitive utilization of controlled quantum adders, wherein the result is accumulated within a dedicated register [9]. Many quantum multipliers [6], [9] belong to this approach. An alternative approach is based on Quantum Fourier Transform (QFT) and, typically, comprises a QFT, controlled phase rotation gates, and an inverse QFT [10]. The multiplier using this approach does not require ancillary qubits but could have a high circuit depth. For example, the out-of-place QFT multiplier has a circuit depth of $O(n^3)$ when tasked with multiplying two $n$-bit integer numbers. Furthermore, there exist other types of quantum multipliers as discussed below. Paper [11] introduces a quantum Karatsuba multiplication, which uses about 10 qubits per bit of input data. Paper [22] proposes a Repeat-Until-Success circuit, which encodes integer numbers as rotation angles within the amplitudes of qubits, subsequently multiplies these rotation angles, and outputs the product as a rotation angle. A common feature shared by these approaches, except for [22], is that they all encode an integer number in binary number system within the computational basis, which requires $n$ qubits to represent an $n$-bit number, where each bit is either 0 or 1.

In this paper, we introduce a new quantum multiplier, called Quantum Multiplier Based on Exponent Adder (QMbead), which uses a new encoding method, called *exponent encoding*. This method represents an integer as a superposition of exponents obtained from its expression as a sum of distinct powers of 2, composing the integer. For example, the integer 14 can be represented as a superposition of $|11\rangle$, $|10\rangle$, and $|01\rangle$, respectively corresponding to the three exponents in this expression: $14 = 2^3 + 2^2 + 2^1$. QMbead takes two integer numbers as input, each employing exponent encoding, and summates them using a quantum adder. The addition results are then measured to derive the product of the two integer numbers. QMbead's key advantage lies in its ability to represent an $n$-bit number using only $\log_2 n$ qubits. Consequently, QMbead can use $2\log_2 n + 1$ qubits to multiply two $n$-bit numbers, aside from an additional $O(n)$ qubits needed for quantum state preparation. Additionally, QMbead has a low circuit depth of $O(\log n)$, lower than existing quantum multipliers. With its efficient use of qubits and low circuit depth, QMbead can multiply very large integer numbers. Note that throughout the paper, when we mention the bit length of an integer, we specifically refer to its bit length when expressed in binary form. For instance, the bit length of 14 is 4, corresponding to its binary form of 1110.

The remainder of the paper is organized as follows. Section II details the QMbead method. Section III conducts a complexity analysis of the QMbead method. Additionally, Section IV presents the numerical results obtained by employing QMbead to multiply both integer and decimal numbers. Finally, in Section V, we draw our conclusions.

## II. Method

In this section, we start with three warm up examples to illustrate the main idea of the QMbead method. Then, we detail the method and provide its Pseudo code. Lastly, we conduct an analysis of the method's complexity.

### Warm Up Examples

In this subsection, we present three illustrative examples of multiplying two numbers. These examples aim to elucidate the motivation and fundamental concept behind the QMbead method.

**Example 1**: multiply two integers, 3 and 5,

$$\begin{aligned} u \times v = 3 \times 5 &= (2^1 + 2^0) \times (2^2 + 2^0) \\ &= 2^1 \times (2^2 + 2^0) + 2^0 \times (2^2 + 2^0) \\ &= (2^{1+2} + 2^{1+0}) + (2^{0+2} + 2^{0+0}) \\ &= 2^3 + 2^1 + 2^2 + 2^0 = 15 \end{aligned} \quad (1)$$

**Example 2**: multiply two integers, 6 and 7,

$$\begin{aligned} u \times v = 6 \times 7 &= (2^2 + 2^1) \times (2^2 + 2^1 + 2^0) \\ &= 2^2 \times (2^2 + 2^1 + 2^0) + 2^1 \times (2^2 + 2^1 + 2^0) \\ &= (2^{2+2} + 2^{2+1} + 2^{2+0}) + (2^{1+2} + 2^{1+1} + 2^{1+0}) \\ &= 2^4 + 2^3 + 2^2 + 2^3 + 2^2 + 2^1 = 42 \end{aligned} \quad (2)$$

These two examples show that we can multiply two integers (in decimal system) by initially decomposing each integer into a sum of distinct powers of 2. Subsequently, applying the distributive property of multiplication, we add the corresponding exponents to obtain the final product expressed in a sum of powers of 2.

As we delve into the QMbead method, we will introduce a unique approach to multiply decimal numbers. Here is an example.

**Example 3**: multiply two decimal numbers, 2.5 and 1.75,

$$\begin{aligned} u \times v &= 2.5 \times 1.75 = (2^1 + 2^{-1}) \times (2^0 + 2^{-1} + 2^{-2}) \\ &= 2^{-1} \times (2^2 + 2^0) \times 2^{-2} \times (2^2 + 2^1 + 2^0) \\ &= 2^{-3} \times (2^2 + 2^0) \times (2^2 + 2^1 + 2^0) \\ &= 2^{-3} \times [(2^2) \times (2^2 + 2^1 + 2^0) + (2^0) \times (2^2 + 2^1 + 2^0)] \\ &= 2^{-3} \times [(2^{2+2} + 2^{2+1} + 2^{2+0}) + (2^{0+2} + 2^{0+1} + 2^{0+0})] \\ &= 2^{-3} \times [2^4 + 2^3 + 2^2 + 2^2 + 2^1 + 2^0] = 2^{-3} \times 35 = 4.375 \end{aligned} \quad (3)$$

Example 3 shows that we can convert a decimal number into the sum of distinct powers of 2, where the exponent can be positive or negative. Then, we multiply each decimal number by a coefficient such that the exponents become non-negative. Then the rest of the process is the same as Examples 1 and 2. In other words, we convert the decimal numbers into integer numbers with appropriate coefficients, followed by multiplying these integer numbers.

*High-level idea of QMbead*: At the core of the QMbead method lies adding the exponents together. We use a quantum adder to implement the addition, leveraging the power of superposition, as elaborated below. Given that each of the number, $u$ and $v$, corresponds to one or more exponents, and the multiplication involves adding each exponent from $u$ to each exponent from $v$, quantum adder is the ideal tool for this task. To facilitate this process, we employ *two quantum states*: $|\psi\rangle$ to represent a superposition of all exponents from $u$ and $|\phi\rangle$ to represent a superposition of all exponents from $v$. We use a *quantum state preparation* method to prepare these two states. Then we use a *quantum adder* to add $|\psi\rangle$ and $|\phi\rangle$ together, where $|\psi\rangle$ and $|\phi\rangle$ can each be either a pure state or a mixed state. In Example 1, $u$ has exponents 1 and 0, which can be represented by $|\psi\rangle$, i.e., $|\psi\rangle = (|1\rangle + |0\rangle)/\sqrt{2}$. Similarly, $v$ has exponents 2 and 0, which can be represented by $|\phi\rangle$, i.e., $|\phi\rangle = (|10\rangle + |00\rangle)/\sqrt{2}$. Our objective in employing the quantum adder is to obtain the desired output: $(|11\rangle + |01\rangle + |10\rangle + |00\rangle)/2$, which is obtained by adding each element in $|\psi\rangle$ with each element in $|\phi\rangle$.

QMbead Method

In this subsection, we detail the QMbead method, along with its Pseudo code.

First, we transform $u$ into a binary representation denoted as $b_{n_u-1} b_{n_u-2} \cdots b_2 b_1 b_0$, which is then converted into a weighted sum of $2^\alpha$, as shown in Eq. (4).

$$u = b_{n_u-1} b_{n_u-2} \cdots b_2 b_1 b_0 = \sum_{\alpha=0}^{n_u-1} b_\alpha 2^\alpha \quad (4)$$

where $b_\alpha \in \{0,1\}$.

Similarly, we transform $v$ into a binary representation denoted as $b_{n_v-1} b_{n_v-2} \cdots b_2 b_1 b_0$, which is then converted into a weighted sum of $2^\beta$, as shown in Eq. (5).

$$v = b_{n_v-1} b_{n_v-2} \cdots b_2 b_1 b_0 = \sum_{\beta=0}^{n_v-1} b_\beta 2^\beta \quad (5)$$

where $b_\beta \in \{0,1\}$.

Then, let $|\psi\rangle$ represent the superposition of all exponents associated with $u$, as expressed in Eq. (6).

$$|\psi\rangle = \left(1/\sqrt{\sum_{\alpha=0}^{n_u-1} b_\alpha^2}\right) \sum_{\alpha=0}^{n_u-1} b_\alpha |\alpha\rangle \quad (6)$$

Similarly, let $|\phi\rangle$ represent the superposition of all exponents associated with $v$, as expressed in Eq. (7).

$$|\phi\rangle = \left(1/\sqrt{\sum_{\beta=0}^{n_v-1} b_\beta^2}\right) \sum_{\beta=0}^{n_v-1} b_\beta |\beta\rangle \quad (7)$$

***Quantum State Preparation***: The preparation of quantum states $|\psi\rangle$ and $|\phi\rangle$ in quantum circuits can be accomplished through a variety of methodologies. While the process of quantum state preparation constitutes an integral component of the QMbead, it falls outside the primary scope of this paper. Consequently, the details of quantum state preparation are omitted herein. Interested readers are directed to Ref. [23], which presents the most contemporary methodology for quantum state preparation.

Then, we use a quantum adder to add states $|\psi\rangle$ and $|\phi\rangle$, with their sum being represented as the output state of the quantum adder, as described in Eq. (8) or Eq. (9), which describe out-of-place and in-place adders, respectively. The details of quantum adders are given in the next subsection.

$$|\psi\rangle|\phi\rangle|0\rangle \to |\psi\rangle|\phi\rangle|\psi+\phi\rangle = |\psi\rangle|\phi\rangle c_0 \sum_{\alpha=0}^{n_u-1} \sum_{\beta=0}^{n_v-1} b_\alpha b_\beta |\alpha+\beta\rangle \tag{8}$$

$$|\psi\rangle|\phi\rangle \to |\psi\rangle|\psi+\phi\rangle = |\psi\rangle c_0 \sum_{\alpha=0}^{n_u-1} \sum_{\beta=0}^{n_v-1} b_\alpha b_\beta |\alpha+\beta\rangle \tag{9}$$

where $c_0 = 1/\sqrt{\sum_{\alpha=0}^{n_u-1} \sum_{\beta=0}^{n_v-1} b_\alpha^2 b_\beta^2}$.

We can reformulate Eqs. (8) and (9) as Eqs. (10) and (11), respectively:

$$|\psi\rangle|\phi\rangle|0\rangle \to |\psi\rangle|\phi\rangle|\psi+\phi\rangle = |\psi\rangle|\phi\rangle c_1 \sum_{\gamma=0}^{n_u+n_v-2} c_\gamma |\gamma\rangle \tag{10}$$

$$|\psi\rangle|\phi\rangle \to |\psi\rangle|\psi+\phi\rangle = |\psi\rangle c_1 \sum_{\gamma=0}^{n_u+n_v-2} c_\gamma |\gamma\rangle \tag{11}$$

where $c_1 = 1/\sqrt{\sum_{\gamma=0}^{n_u+n_v-2} c_\gamma^2}$.

By measuring the output of the quantum adder, i.e., state $|\psi+\phi\rangle$, we obtain state $|\gamma\rangle$ with a probability of $p_\gamma$. If only one state, denoted as $|\gamma\rangle$, is measured, we obtain the *product* as

$$u \times v = 2^\gamma \tag{12}$$

On the other hand, if multiple states are measured, denoted as $|\gamma_0\rangle, |\gamma_1\rangle, \cdots, |\gamma_{n_\gamma}\rangle$, with the probabilities of $p_0, p_1, \cdots, p_{n_\gamma}$, respectively, then we can obtain the *product* as

$$u \times v = \sum_{i=0}^{n_\gamma} (p_i/p_{\min}) \, 2^{\gamma_i} \tag{13}$$

where $p_{\min}$ represents the minimum value among all probabilities $p_1, p_2, \cdots, p_{n_\gamma}$, and $\gamma_i$ is converted into its decimal form.

***Number of Shots***: Here we discuss how to set the number of shots required to estimate the $p_i$ and $p_{\min}$ in Eq. (13). In quantum computing, we perform measurements multiple times, referred to as *shots*, to observe each possible output and estimate its probability based on the frequency of measurement occurrences. To simplify expression, we refer to $\gamma_i$ as the *exponent of the product*. The number of exponents of the product with non-zero probability ($p_i$) affects the number of shots required to measure both low-probability and high-probability exponents of the product. However, the number of exponents with non-zero probability is unknown until we determine the product value, although we can set it as the product of the number of $b_\alpha$ that is equal to 1 and the number of $b_\beta$ that is equal to 1. Alternatively, we can set the number of shots based on the number of qubits needed to represent the product. Since we need to round each $p_i/p_{\min}$ in Eq. (13) to the nearest integer and measurement outcomes are probabilistic, a sufficiently large number of shots is essential to accurately calculate $p_i/p_{\min}$. We can set the number of shots to be $C_0 \times 2^{1+\lceil \log_2 n_m \rceil}$ such that we will obtain the lowest-probability exponent of product around $C_0$ times, where $C_0$ denotes a coefficient (we can choose $C_0$ within the range of 1000~10000), $1 + \lceil \log_2 n_m \rceil$ represents the number of qubits required to represent the exponents of the product, and $n_m$ is the larger of the bit lengths of the two multiplicands, i.e., $n_m = \max(n_u, n_v)$.

To provide an intuitive understanding of how the number of shots needed scales with the bit length $n_m$, we plot the function $2000 \times 2^{1+\lceil \log_2 n_m \rceil}$ in Fig. S1. The figure shows a piecewise linear growth in the number of shots with respect to the bit length of the larger multiplicand.

The Pseudo code for QMbead is given in Algorithm 1.

---

**Algorithm 1: Pseudo code for QMbead**

**Input**: integer numbers $u$ and $v$.

**Output**: the product of the two numbers, denoted as $u \times v$.

1. Convert $u$ into a sum of distinct powers of 2 using Eq. (4) and $v$ into a similar representation using Eq. (5).
2. Obtain states $|\psi\rangle$ and $|\phi\rangle$, each representing a superposition of their respective exponents, using Eqs. (6) and (7), respectively. Prepare the quantum states $|\psi\rangle$ and $|\phi\rangle$ in quantum circuits using a quantum state preparation method [23].
3. Utilize a quantum adder to add states $|\psi\rangle$ and $|\phi\rangle$, resulting in the output state $|\psi + \phi\rangle$, as described in Eqs. (10) and (11), for out-of-place and in-place adders, respectively.
4. Measure the output state of the quantum adder, i.e., $|\psi + \phi\rangle$, to calculate the product of $u$ and $v$. If only one state is measured, the product is determined by Eq. (12). If multiple states are measured, the product is computed using Eq. (13).

---

Quantum Adder

Here we provide two versions of QFT-based quantum adders to complete step 3 in Algorithm 1. The first version is an in-place quantum adder, as shown in Fig. 1. The input and output of this circuit can be written as $|\phi\rangle_B |0\rangle_C |\psi\rangle_A$ and $|\phi\rangle_B |\phi + \psi\rangle_{CA}$.

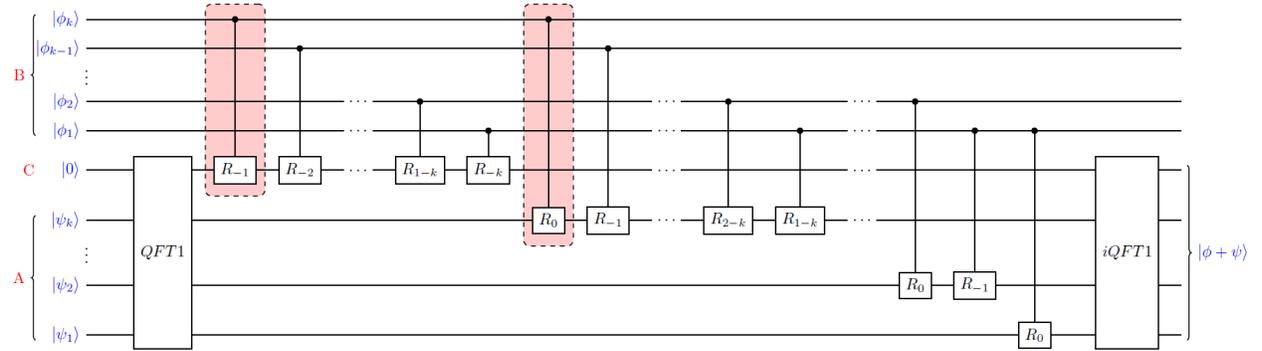

Fig. 1. The in-place quantum adder based on QFT (version 1). It takes states $|\phi\rangle$ and $|\psi\rangle$ as input and outputs their sum $|\phi + \psi\rangle$. It consists of three registers, B, C, and A (indicated by the red letters on the leftmost side). Register C contains only one qubit, called *carry qubit*. Register A comprises $k$ qubits, used for representing state $|\psi\rangle$. Register B is for state $|\phi\rangle$ and has $k$ qubits, although it can have fewer qubits if needed. For example, if $|\phi\rangle$ has only $k-1$ qubits, then the top wire, $|\phi_k\rangle$, and the two controlled rotation gates in dashed red box can be deleted. The '*QFT1*' and '*iQFT1*' denote the QFT without swap gates and its inverse, respectively.

In both Figs. 1 and 2, the symbol $R_x$ represents the gate $\begin{bmatrix} 1 & 0 \\ 0 & e^{i\pi 2^x} \end{bmatrix}$. Note that $R_z(\theta) = e^{-i\theta/2} \begin{bmatrix} 1 & 0 \\ 0 & e^{i\theta} \end{bmatrix}$. Then we can realize the $R_x$ gate as an $R_z(\theta)$ gate with $\theta = \pi 2^x$, up to a global phase difference.

Note that the carry qubit (with the red 'C' on the leftmost side) in the circuit given in Fig. 1 does not have a controlled $R_0$ gate, while all the $k$ wires for state $|\psi\rangle$ has a controlled $R_0$ gate. Here state $|\phi\rangle$ remains unchanged (i.e., the input and output of register B are the same), while state $|\psi\rangle$ can change during the operation of this circuit. This quantum adder is like, but slightly different from, the one in [24]. Importantly, the circuit in Fig. 1 does not suffer from overflow issues, unlike the latter, and it allows the number of qubits in register B to be equal to or less than that in register A.

If we need to maintain both the states $|\phi\rangle$ and $|\psi\rangle$ unchanged throughout the computation, we can use the out-of-place quantum adder depicted in Fig. 2, where $|s_i\rangle=|0\rangle$ for all $i$ in $[1,k+1]$. The input and output of this circuit can be written as $|\phi\rangle|\psi\rangle|0\rangle$ and $|\phi\rangle|\psi\rangle|\psi+\phi\rangle$, respectively. This adder requires more qubits and has a higher circuit depth compared to the one shown in Fig. 1.

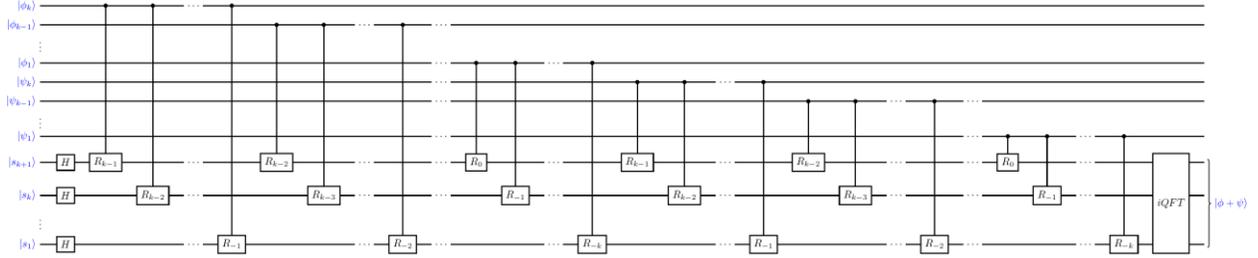

Fig. 2. The out-of-place quantum adder based on QFT (version 2). It takes states $|\phi\rangle$ and $|\psi\rangle$ as input and produces their sum $|\phi+\psi\rangle$ as output. This adder ensures that the states $|\phi\rangle$ and $|\psi\rangle$ remain unchanged throughout the computation.

### III. Complexity Analysis

In this section, we provide the complexity analysis, including the number of qubits and the circuit depth, for the QMbead using four different adders: the in-place and out-of-place quantum adders based on QFT, as shown in Figs. 1 and 2, respectively, and the in-place and out-of-place logarithmic-depth quantum carry lookahead adder (QCLA) discussed in [4].

As shown in Eq. (4), integer $u$ consists of $n_u$ binary bits and it is expressed as a superposition of states $|\alpha\rangle$ according to Eq. (6). The maximum value of $\alpha$ is $n_u - 1$, which can be represented by $\lceil \log_2 n_u \rceil$ qubits, where the notation $\lceil x \rceil$ denotes the smallest integer greater than or equal to $x$. Similarly, we use $\lceil \log_2 n_v \rceil$ qubits to represent the components associated with the integer $v$. For instance, we can

- represent a 9-bit number, 384 (equal to $2^8 + 2^7$), as a four-qubit state: $(|1000\rangle + |0111\rangle)/\sqrt{2}$,
- represent a 21-bit number, 1835008 (equal to $2^{20} + 2^{19} + 2^{18}$), as a 5-qubit state: $(|10100\rangle + |10011\rangle + |10010\rangle)/\sqrt{3}$, and
- represent a 1024-bit number, $2^{1023} + 2^{1022}$ (approximating to $1.35 \times 10^{308}$), as a 10-qubit state $(|1\,111\,111\,111\rangle + |1\,111\,111\,110\rangle)/\sqrt{2}$.

For simplicity, we assume $n_u = n_v = n$ in the rest of the paper, meaning integers $u$ and $v$, when expressed in binary form, have an equal number of binary bits. However, it is worth noting that the circuits shown in Figs. 1 and 2 can handle two integers with different bit lengths, represented by different numbers of qubits.

Quantum Adder:

For the QMbead using the quantum adder given in Fig. 1, it needs $2\lceil \log_2 n \rceil + 1$ qubits for multiplying integers $u$ and $v$. In the case of the QMbead utilizing the quantum adder given in Fig. 2, it needs $3\lceil \log_2 n \rceil + 1$ qubits for multiplying integers $u$ and $v$. Note that in the following three paragraphs, $k = \lceil \log_2 n \rceil$.

In Fig. 1, the (k+1)-qubit QFT without swap gates and its inverse each has a circuit depth of $(k+2)(k+1)/2$ and the depth of the controlled rotation gates is $(k+3)k/2$. If we consider controlled rotation gates on different qubits can be applied in parallel [24], the depth of controlled rotation gates is $k$. Then we can express the total depth of the circuit given in Fig. 1 as $k + 2 \times (k+2)(k+1)/2 = k^2 + 4k + 2$.

Now we consider Fig. 2. The circuit depth of the Hadamard gates is 1. The inverse of the (k+1)-qubit QFT, denoted as *iQFT*, has a circuit depth of $\frac{(k+2)(k+1)}{2} + \lfloor \frac{k+1}{2} \rfloor$, where the notation $\lfloor x \rfloor$ denotes the largest integer smaller than or equal to $x$. The depth of controlled rotation gates is $2k(k+1)$. If we consider controlled rotation gates on different qubits can be applied in parallel, the depth of the *iQFT* reduces to $(k+2)(k+1)/2 + 1$ and the depth of controlled rotation gates becomes $2(k+1)$. Then we can express the total depth of the circuit given in Fig. 2 as
$$1 + (k+2)(k+1)/2 + 1 + 2(k+1) = (k^2 + 7k)/2 + 4.$$

In Section II, we demonstrated two quantum adders based on QFT. However, it is vital to note that alternative quantum adders, such as the logarithmic-depth quantum carry lookahead adder (QCLA) discussed in [4], can be implemented in QMbead to further reduce the circuit depth. According to [4], to add two $n$-bit numbers, the QCLA out-of-place adder needs $4n + 1 - \omega(n) - \lfloor \log_2 n \rfloor$ qubits with a circuit depth of $2\log_2 n + O(1)$, where $\omega(n)$ denotes the number of ones in the binary expansion of $n$. The QCLA in-place adder needs $4n - 1 - \omega(n) - \lfloor \log_2 n \rfloor$ qubits with a circuit depth of $4\log_2 n + O(1)$. In QMbead, we need to add two $\lceil \log_2 n \rceil$-bit numbers. Then the QCLA out-of-place adder for the QMbead needs $4\lceil \log_2 n \rceil + 1 - \omega(\lceil \log_2 n \rceil) - \lfloor \log_2 \lceil \log_2 n \rceil \rfloor$ qubits with a circuit depth of $2 \log_2 \lceil \log_2 n \rceil + O(1)$, and the QCLA in-place adder for the QMbead needs $4\lceil \log_2 n \rceil - 1 - \omega(\lceil \log_2 n \rceil) - \lfloor \log_2 \lceil \log_2 n \rceil \rfloor$ qubits with a circuit depth of $4\log_2 \lceil \log_2 n \rceil + O(1)$. For the ease of comparison, Table I summarizes the number of qubits and the circuit depth of the QMbead using different types of quantum adders, where IP and OOP indicate in-place and out-of-place adders, respectively. The in-place and out-of-place QFT adders for QMbead are given in Figs. 1 and 2, respectively. The last two rows of Table I are for the QMbead using the QCLA as the adder.

Table I. Qubit count and circuit depth for QMbead using quantum adders from Figs. 1 and 2.

| QMbead | # of qubits[†] | Circuit depth[*] |
|---|---|---|
| QFT, IP | $2\lceil \log_2 n \rceil + 1$ | $\lceil \log_2 n \rceil^2 + 4\lceil \log_2 n \rceil + 2$ |
| QFT, OOP | $3\lceil \log_2 n \rceil + 1$ | $(\lceil \log_2 n \rceil^2 + 7\lceil \log_2 n \rceil)/2 + 4$ |
| QCLA, IP | $4\lceil \log_2 n \rceil - 1 - \omega(\lceil \log_2 n \rceil) - \lfloor \log_2 \lceil \log_2 n \rceil \rfloor$ | $4 \log_2 \lceil \log_2 n \rceil + O(1)$ |
| QCLA, OOP | $4\lceil \log_2 n \rceil + 1 - \omega(\lceil \log_2 n \rceil) - \lfloor \log_2 \lceil \log_2 n \rceil \rfloor$ | $2 \log_2 \lceil \log_2 n \rceil + O(1)$ |

[†] The qubits used for state preparation are $O(n)$ as explained below but not considered in the table.
[*] Parallel application of controlled rotation gates is considered for circuit depth calculation. We assume $n_u = n_v = n$.

Quantum State Preparation and Circuit Depth Complexity:

Now let us consider the complexity of preparing the initial superposition states which is given in Eqs. (6) and (7). Different methods for preparing a quantum state are available. In this paper, when we calculate the number of qubits, the circuit depth, and the time and gate complexities of quantum state preparation, we assume the best quantum state preparation method in the literature, i.e., [23], is used in QMbead. According to Theorem 1 of paper [25], with only single- and two-qubit gates, an arbitrary $n$-qubit quantum state can be deterministically prepared with a circuit depth $\Theta(n)$ and $O(2^n)$ ancillary qubits. The method given in [23] also involves a one-time classical preprocessing to calculate phase angles needed in the method, which take time $O(2^n)$ by sequential calculations. In QMbead, we need to add two $\lceil \log_2 n \rceil$-bit numbers. As a result, the state preparation for QMbead needs a circuit with a depth of $\Theta(\log n)$ and $O(n)$ ancillary qubits and needs time $O(n)$ for the classical preprocessing. Consequently, the circuit depth of the entire QMbead, including both state preparation and the QCLA adder, is $O(\log n)$.

Classical Part of QMbead:

Now let us delve into the complexity analysis of the classical component within QMbead. We assume that both multiplicands are in binary form, and our objective is to obtain the product in binary form as well. The multiplicand in binary form can be directly utilized by the quantum state preparation algorithm, eliminating the need for classical computation. Regarding the output, the measured state $|\gamma_i\rangle$ is represented in binary form and has $O(\log n)$ bits. Converting this state into its decimal form, called the *binary-to-decimal conversion*, carries a complexity of $O(\log n)$. To reduce the computational complexity, we can implement pre-prepared list that links each possible state $|\gamma_i\rangle$, which is in its binary form, with its corresponding decimal number. In this optimized scenario, the complexity of the binary-to-decimal conversion becomes $O(1)$. The number of states $|\gamma_i\rangle$ is denoted as $n_\gamma$, and it can be expressed as $O(2n)$. Consequently, the classical component of the calculation involved in the conversion from measurement results to the product in binary form holds a complexity of $O(n)$.

Time Complexity and Gate Complexity:

As mentioned above, we set the number of measurement shots to be $C_0 \times 2^{1+\lceil \log_2 n \rceil}$, which can be represented as $O(n)$. By multiplying this with the circuit depth (detailed two paragraphs above), we express the *time complexity* for the quantum part of QMbead, using QCLA as the quantum adder, as $O(n \log n)$.

The classical part of QMbead has lower complexity as described in the previous paragraph. The classical preprocessing of the method given in [23] has a time complexity of $O(n)$ as mentioned in the 'Quantum State Preparation' subsection, which is also lower than the quantum part $O(n \log n)$. Therefore, the total *time complexity* of the QMbead is $O(n \log n)$.

The quantum circuits of the QMbead include both the quantum state preparation and the quantum adder. According to Algorithms 1-3 given in the [23], the gate complexity of the state preparation method is $O(n)$. The gate complexity of the in-place quantum adder based on QFT for adding two $\lceil \log_2 n \rceil$-bit numbers is $O(\log^2 n)$. According to [4], the gate complexity of the in-place and out-of-place QCLA for adding two $\lceil \log_2 n \rceil$-bit numbers is $8 \log n - O(\log_2 \lceil \log_2 n \rceil)$ and $16 \log n - O(\log_2 \lceil \log_2 n \rceil)$, respectively. The gate complexity of each of these adders is lower than that of the state preparation method. Therefore, the *gate complexity* of the QMbead is $O(n)$, determined by the state preparation stage.

Complexity Comparison:

Here we compare the complexity of the QMbead with the two best quantum multipliers. The quantum Karatsuba multiplication [11] has $O(n^{1.58})$ gate complexity. The out-of-place QFT multiplier, which is a widely used quantum multiplier, has a circuit depth of $O(n^3)$. As described above, the circuit depth of the entire QMbead is $O(\log n)$, and the gate complexity of the QMbead is $O(n)$. Therefore, QMbead has lower complexity compared to existing quantum multiplication algorithms.

The fastest classical algorithm for multiplying two *n*-digit numbers is the Harvey-Hoeven algorithm, proposed in 2019 [26]. Its time complexity is $O(n \log n)$. However, it is essential to note that this algorithm only works for ludicrously large numbers but not practical in its current form, as acknowledged by one of the authors of the paper[1]. Another notable classical algorithm is the Schönhage–Strassen algorithm [27], which has a time complexity of $O(n \log n \log \log n)$. Therefore, QMbead shares the same time complexity as the fastest classical multiplication algorithm, the Harvey-Hoeven algorithm. However, QMbead is valid for both small and larger numbers, while the Harvey-Hoeven algorithm is specifically suitable for ludicrously large numbers. This versatility gives QMbead a distinct *advantage* over the fastest classical multiplication algorithm.

---

[1] https://theconversation.com/weve-found-a-quicker-way-to-multiply-really-big-numbers-114923

## IV. Numerical Results

To verify its effectiveness, we executed the QMbead on Pennylane's qubit device [28], performing multiplications for 13 pairs of integer numbers and 3 pairs of decimal numbers. The obtained results are presented in Table II. The correct product values are given in the 4[th] column. The number of shots used in the measurement is given in the 5[th] column. Table II shows that as the bit length of the product (the 6[th] column) increases from 4 to 273, the number of qubits (the 7[th] column) used by QMbead, utilizing the quantum adder from Fig. 1, grows slowly. We plot their relationship in Fig. 3, showcasing a logarithmic correlation, i.e., the number of qubits required by QMbead is the logarithm of the bit length of the product, where the ancillary qubits used in the state preparation are not considered.

The exponents associated with the product are represented by the output state of the QMbead in the quantum circuit. If we can obtain the theoretical probability of each exponent, using Eq. (13) ensures the calculation of the accurate value every time. However, accessing information from a quantum circuit is solely achievable through measurements. As mentioned above, calculating a precise value for $p_i/p_{\min}$ requires a sufficiently large number of measurements.

The number of shots specified in the 5[th] column of Table II is determined via a trial-and-error approach. We initially set it to $1 \times 10^5$. Should the resulting product be inaccurate, we iteratively increase the number to the subsequent minimum value—a multiple of either $1 \times 10^z$ or $5 \times 10^z$, where $z$ is an integer. This iteration continues until we achieve the correct product once.

To assess the QMbead's stability with the designated number of shots in each row of the table, we independently execute the QMbead for each case 200 times. The last column of Table II provides the frequency of obtaining the correct product out of these 200 times. In most instances (13 out of 16 cases), we achieve the correct product in over 195 out of 200 times. This high success rate suggests that the number of shots for these 13 cases is adequately large for the QMbead to reliably produce the correct product.

In the remaining three cases (id=7, 12, and 13), the occurrences of obtaining the correct product are 173, 163, and 121, respectively. Although these frequencies are lower than the other 13 cases, indicating that the number of shots might not be sufficiently large, they are still sufficient to obtain the correct product, with the most frequently recurring value representing the correct product. Furthermore, we can expect that increasing the number of shots in these three cases can yield a higher success rate in obtaining the correct product.

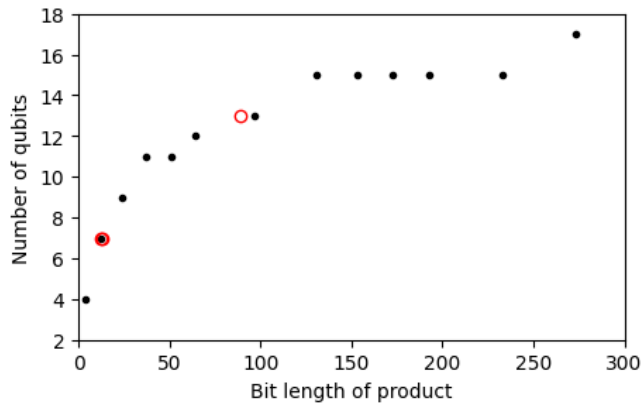

Fig. 3. Scatter plot of the relationship between the number of qubits used by QMbead with the quantum adder given in Fig. 1 and the bit length of the product. The black dots correspond to the integer cases (the first 13 cases in Table II), while the red circles represent the decimal cases (the last 3 cases in Table II). Note that the ancillary qubits used in the state preparation are not considered.

Table II. The multiplicands, $u$ and $v$, and the product obtained by QMbead with the quantum adder given in Fig. 1. The number of shots, the bit length of the product, and the total number of qubits used in the QMbead are provided in the last three columns. In the last three columns, the BLP, #Q, and #C respectively represent the bit length of product, the number of qubits, and the number of times obtaining the correct product out of 200 independent runs.

| id | Multiplicand $u$ | Multiplicand $v$ | Product ($u \times v$) | # of shots | BLP | #Q[†] | #C |
|---|---|---|---|---|---|---|---|
| 1 | 3 | 5 | 15 | $1 \times 10^5$ | 4 | 4 | 200 |
| 2 | 33 | 100 | 3300 | $1 \times 10^5$ | 12 | 7 | 200 |
| 3 | 2345 | 5678 | 13314910 | $1 \times 10^5$ | 24 | 9 | 200 |
| 4 | 234501 | 567801 | 133149902301 | $1 \times 10^5$ | 37 | 11 | 197 |
| 5 | 23450101 | 56780101 | 1331499103240201 | $1 \times 10^6$ | 51 | 11 | 200 |
| 6 | 2345010101 | 5678010101 | 13314991040425030201 | $1 \times 10^6$ | 64 | 12 | 196 |
| 7 | 8978923748987 | 8984957438475849 | 80675247727968202502337714963 | $5 \times 10^6$ | 97 | 13 | 173 |
| 8 | 24587098456973459873 | 93847898723487384738 | 2307447525894458211799840656192155618274 | $5 \times 10^7$ | 131 | 15 | 200 |
| 9 | 98734587398457983758948 | 87234593847548978394754 | 8613071630409809722603055112891858675923758792 | $5 \times 10^7$ | 153 | 15 | 199 |
| 10 | 873498539875897899837437878 | 945430854908947584788947548 | 8258324713165875922142248753007708733319714270423144 | $5 \times 10^7$ | 173 | 15 | 198 |
| 11 | 98734574983957438978459843787 | 91398475934873485748398475397 | 9024189675611195666675703027302141983724295908377582808439 | $1 \times 10^8$ | 193 | 15 | 200 |
| 12 | 87892734987329734982798374239878729 | 99787498783927389473829348739287348 | 8770596185664218247269027408953463166804991960126150471650153404020692 | $1 \times 10^8$ | 233 | 15 | 163 |
| 13 | 98789236479326873476287376473627847267623 | 92934837483278492837489283478928374829373 | 9180961637303410170533798160931075913254844659074625888400424445374174806892290379 | $1 \times 10^8$ | 273 | 17 | 121 |
| 14 | 0.567 | 0.0004 | 0.0002268 | $1 \times 10^5$ | 12[*] | 7 | 200 |
| 15 | 2.5 | 1.75 | 4.375 | $1 \times 10^5$ | 13[*] | 7 | 200 |
| 16 | 136872.345502 | 2343651.74543455 | 320781111437.483078727894100 | $5 \times 10^6$ | 89[*] | 13 | 197 |

[†]The qubits used for state preparation are not considered here.
[*]The bit lengths in these three rows correspond to the bit length of the scaled product, which is an integer.

Table III. Success rate of the QMbead runs with varied numbers of shots, for the case with 'id' equal to 7 in Table II. Each setting of shots reflects 200 independent runs. The #C denotes the frequency of correct product occurrence out of 200 runs.

| # of shots | $1 \times 10^6$ | $2.5 \times 10^6$ | $5 \times 10^6$ | $7.5 \times 10^6$ | $10 \times 10^6$ |
|---|---|---|---|---|---|
| #C | 13 | 76 | 173 | 199 | 197 |
| Success rate | 0.065 | 0.38 | 0.865 | 0.995 | 0.985 |

To explore the impact of the number of shots on the success rate of the QMbead, we take the case with 'id' equal to 7 in Table II as an example. The *success rate* is defined as the number of times that the QMbead obtains the correct product divided by the total number of runs (which is 200 in this paper). We varied the number of shots across five different values. For each of the five settings, we conducted 200 independent runs of the QMbead, with the frequency of obtaining the correct product and the success rate detailed in Table III.

The results in Table III indicate that with the number of shots set at $1 \times 10^6$ and $2.5 \times 10^6$, the frequency of obtaining the correct product is notably low, at 13 and 76 out of 200 times, respectively. However, when the number of shots increases to $7.5 \times 10^6$ and $10 \times 10^6$, the success rate substantially rises to 199 and 197 out of 200 times, respectively. This suggests that setting the number of shots to $7.5 \times 10^6$ in this case (id=7) proves to be sufficiently large. Given that the measurement outcomes are inherently probabilistic, we consider the success rates in the last two columns are at the same level, being nearly 100%, rather than exhibiting a slight decrease with an increase in the number of shots. Thus, Table III demonstrates a trend: as the number of shots increases, so does the success rate, aligning with our expectations.

Now we present the measurement results for three cases: 3×5, 33×100, and 2345×5678, shown as bar plots in Figs. 4-6, respectively. The left panel of each figure (Figs. 4-6) provides the number of counts for each measured state, denoted by the digits at the bottom of each bar. The right panel of each of figure (Figs. 4-6) shows the coefficient $p_i/p_{\min}$ in Eq. (13). These coefficients have been rounded to the nearest integer for each exponent of the product, i.e., $\gamma_i$ in Eq. (13), where the exponent of the product is represented by the digits in binary form at the bottom of each bar. Note that both panels share the same x-axis labels, despite their distinct meanings. The former indicates measured states, while the latter represents exponents. Each measured state is associated with a specific exponent. For instance, the x-axis labels of the four bars in Fig. 4 are 000, 001, 010, and 011, representing the states $|000\rangle$, $|001\rangle$, $|010\rangle$, $|011\rangle$, and the corresponding exponents in binary form, respectively.

We calculate the probability $p_i$ in Eq. (13) by dividing the number of counts by the total number of shots. We illustrate how to derive the correct product from the right panel of each figure (Figs. 4-6), as detailed in each figure caption, where the exponents in binary form are converted to decimal form.

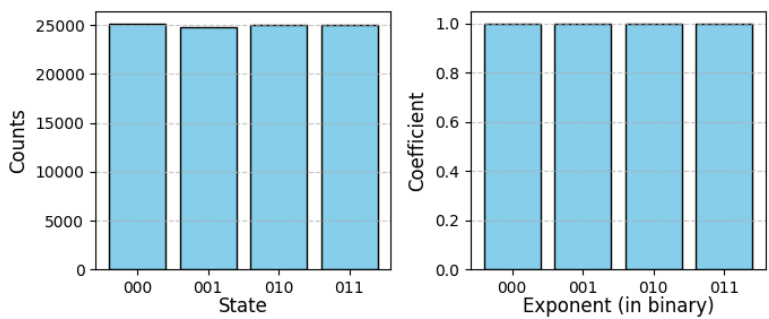

Fig. 4. The measurement counts (left panel) and the rounded coefficient $p_i/p_{\min}$ in Eq. (13) (right panel) obtained from the output of QMbead for calculating 3×5. According to Eq. (13), the product is determined from the right panel as $1 \times 2^0 + 1 \times 2^1 + 1 \times 2^2 + 1 \times 2^3$, which is equal to 15, where the exponents are obtained by converting the binary digits to their respective decimal values.

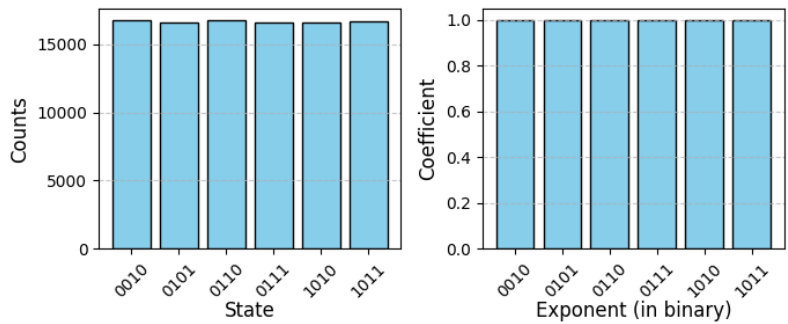

Fig. 5. The measurement counts (left panel) and the rounded coefficient $p_i/p_{\min}$ (right panel) obtained from the output of QMbead for calculating 33×100. The digits in the x-axis should be read from bottom left to top right. According to Eq. (13), the product is determined from the right panel as $1 \times 2^2 + 1 \times 2^5 + 1 \times 2^6 + 1 \times 2^7 + 1 \times 2^{10} + 1 \times 2^{11}$, which is equal to 3300.

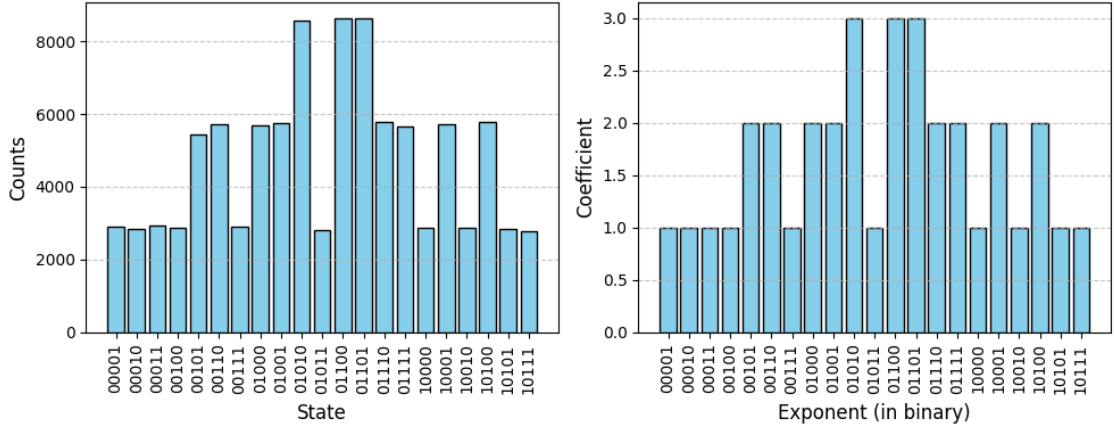

Fig. 6. The measurement counts (left panel) and the rounded coefficient $p_i/p_{\min}$ (right panel) obtained from the output of QMbead for calculating 2345×5678. The digits in the x-axis should be read from bottom to top. According to Eq. (13), the product is determined from the right panel as $2^1 + 2^2 + 2^3 + 2^4 + 2 \times 2^5 + 2 \times 2^6 + 2^7 + 2 \times 2^8 + 2 \times 2^9 + 3 \times 2^{10} + 2^{11} + 3 \times 2^{12} + 3 \times 2^{13} + 2 \times 2^{14} + 2 \times 2^{15} + 2^{16} + 2 \times 2^{17} + 2^{18} + 2 \times 2^{20} + 2^{21} + 2^{23}$, which is equal to 13314910.

## V. Conclusion

This paper introduces the QMbead method designed for efficient multiplication of both integer and decimal numbers. The quantum part of the QMbead method consists of a state preparation method and a quantum adder. When analyzing the complexity of the QMbead, we have considered two types of adders: the QFT-based quantum adders and QCLA. To achieve precise product calculations, the number of measurement shots must be sufficiently large, scaling linearly with the bit length of the product. In our implementation of QMbead, two versions of QFT-based quantum adders are employed. QMbead has demonstrated effectiveness in the multiplication of large integers (up to 273 bits in our numerical study) or decimal numbers with many bits, showcasing potential applications in large integer factorization challenges.

When using QCLA as the adder, QMbead requires $O(\log n)$ qubits to accurately multiply two *n*-bit numbers and its circuit depth has a complexity of $O(\log_2[\log_2 n])$, in addition to $O(n)$ ancillary qubits and $O(\log n)$ circuit depth used in state preparation stage. The *gate complexity* of the QMbead is $O(n)$. The *time complexity* of the QMbead, using QCLA as the adder, is $O(n \log n)$, equivalent to the fastest classical multiplication algorithm. However, QMbead holds the advantage of applicability to both large and small numbers, while the classical algorithm is only practical for extremely large numbers. The QMbead has also shown lower complexities than existing quantum multipliers such as quantum Karatsuba multiplication and QFT multiplier.

## Code Availability

The code for the QMbead and the results given in Section IV are available at https://github.com/zhanjun-peng/QMbead.

## Acknowledgments

We acknowledge the support from the NSF ERI program, under award number 2138702. This work used the Delta system at the National Center for Supercomputing Applications through allocations CIS220136 and ELEC220008 from the Advanced Cyberinfrastructure Coordination Ecosystem: Services & Support (ACCESS) program, which is supported by National Science Foundation grants #2138259, #2138286, #2138307, #2137603, and #2138296. The author of this paper thanks Dr. Markus Grassl for useful discussions.

## Appendix

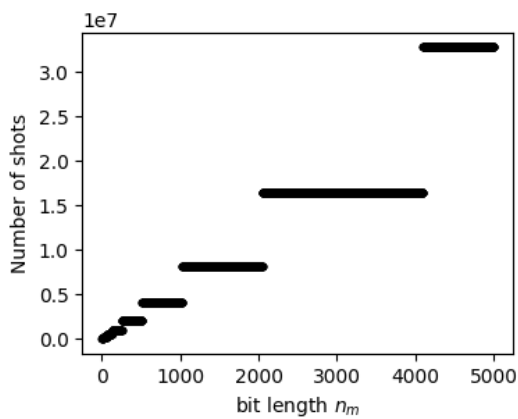

Fig. S1. The scatter plot showing the relationship between the number of shots needs and the bit length, $n_m$, of the larger multiplicand, i.e., a plot of the function $2000\times 2^{1+\lceil \log_2 n_m \rceil}$.